\documentclass[prd,preprint,linenumbers,amssymb,amsmath,aps,nofootinbib,superscriptaddress]{revtex4}

\usepackage[utf8]{inputenc} 
\usepackage[T1]{fontenc}    
\usepackage{booktabs}       
\usepackage{amsfonts}       
\usepackage{nicefrac}       
\usepackage{microtype}      
\usepackage{graphicx}
\usepackage{natbib}
\usepackage{doi}
\usepackage{mathrsfs}
\usepackage{caption}
\usepackage{overpic}
\usepackage{adjustbox}
\usepackage{graphicx}
\usepackage{color}
\usepackage{subcaption} 
\usepackage{float} 

\begin{document}
\title{New Constraints on Cosmic-Ray Boosted Sub-GeV Dark Matter via Light Mediators}

\author{Yang Yu}
\affiliation{Key Laboratory of Dark Matter and Space Astronomy, Purple Mountain Observatory, Chinese Academy of Sciences, Nanjing 210023, China}
\affiliation{School of Astronomy and Space Science, University of Science and Technology of China, Hefei 230026, China}

\author{Guan-Sen Wang}
\affiliation{Key Laboratory of Dark Matter and Space Astronomy, Purple Mountain Observatory, Chinese Academy of Sciences, Nanjing 210023, China}
\affiliation{School of Astronomy and Space Science, University of Science and Technology of China, Hefei 230026, China}

\author{Bo Zhang}
\affiliation{Key Laboratory of Dark Matter and Space Astronomy, Purple Mountain Observatory, Chinese Academy of Sciences, Nanjing 210023, China}
\affiliation{School of Astronomy and Space Science, University of Science and Technology of China, Hefei 230026, China}

\author{Tian-Peng Tang}
\email{tangtp@pmo.ac.cn}
\affiliation{Key Laboratory of Dark Matter and Space Astronomy, Purple Mountain Observatory, Chinese Academy of Sciences, Nanjing 210023, China}
\author{Bing-Yu Su}
\email{bysu@pmo.ac.cn}
\affiliation{Key Laboratory of Dark Matter and Space Astronomy, Purple Mountain Observatory, Chinese Academy of Sciences, Nanjing 210023, China}
\author{Lei~Feng}
\email{fenglei@pmo.ac.cn}
\affiliation{Key Laboratory of Dark Matter and Space Astronomy, Purple Mountain Observatory, Chinese Academy of Sciences, Nanjing 210023, China}
\affiliation{School of Astronomy and Space Science, University of Science and Technology of China, Hefei 230026, China}
\affiliation{Joint Center for Particle, Nuclear Physics and Cosmology,  Nanjing University -- Purple Mountain Observatory,  Nanjing  210093, China}

\begin{abstract}

Traditional direct detection experiments lack the sensitivity to probe the sub-GeV dark matter(DM), primarily due to the low energy of the expected nuclear recoils.
In this work, we investigate cosmic-ray (CR) upscattering as a mechanism to accelerate DM particles to detectable velocities in underground experiments. 
By analyzing four models of DM–nucleon interactions—namely scalar, vector, pseudoscalar, and axial-vector mediators—we derive constraints on the coupling parameters using data from the LZ, XENON, and Borexino experiments, covering mediator mass from $10^{-6}$ to $1$ GeV.
As the mediator mass varies, the shift in dominance between momentum transfer and mediator mass leads to a turnover in the constraints around $10^{-2}$--$10^{-3}~\mathrm{GeV}$.
Our results extend the reach of direct detection into the sub-GeV window and clarify the critical role of momentum dependence in light-mediator scenarios.

\end{abstract}

\maketitle

\section{Introduction}
The existence of dark matter (DM), accounting for approximately 85\% of the matter content in the universe~\cite{Zwicky:1933gu,Planck:2018vyg}, stands as one of the most compelling and unresolved puzzles in modern physics.
While its gravitational imprint---from galactic rotation curves~\cite{Rubin:1980zd} to the formation of large-scale~\cite{Clowe:2006eq}---provides irrefutable evidence, the particle nature of DM remains elusive.  

Among the leading DM candidates, weakly interacting massive particles (WIMPs)~\cite{Bertone:2004pz} have dominated direct detection efforts for decades. These experiments aim to identify nuclear recoils induced by DM scattering ultra-low-background detectors deep underground~\cite{Lewin:1995rx}. 
Key experiments in this field include XENON~\cite{XENON:2019zpr,XENON:2022ltv,XENON:2023cxc}, LUX-ZEPLIN (LZ)~\cite{LZ:2022lsv,LZ:2023poo,LZ:2024zvo}, PandaX~\cite{PandaX-II:2020oim,PandaX:2024qfu,PandaX:2025rrz}, DEAP-360~\cite{DEAP:2019yzn}, and DarkSide-50~\cite{DarkSide:2018kuk}.
However, conventional searches confront a fundamental sensitivity gap in the sub-GeV mass regime.
The keV-scale energy thresholds typical of nuclear recoil detectors are often too high to register the faint signals induced by non-relativistic Galactic DM particles, whose velocities are only $\sim 220\rm{km/s}$ within the standard halo model~\cite{Knapen:2017xzo, Essig:2011nj}.
Consequently, light DM particles evade detection due to insufficient momentum transfer, leaving a vast and theoretically motivated parameter space largely unexplored.

To bridge this sensitivity gap, significant progress has been made on both experimental~\cite{SuperCDMS:2018mne,DarkSide:2022dhx,DarkSide-50:2022qzh,DAMIC:2016lrs,NEWS-G:2017pxg,CRESST:2019jnq,CDEX:2019hzn,EDELWEISS:2019vjv,EDELWEISS:2022ktt} and theoretical fronts~\cite{Essig:2017kqs,Chao:2021liw,Wang:2025jhy,Zhang:2024qof,Guo:2023kqt,Tang:2025vqf,Liang:2024xcx,Su:2023zgr,Chen:2024njd,Zhang:2024mmg,Wang:2025tdx,Yu:2024xsy}.
One particularly compelling theoretical proposal is the cosmic-ray boosted dark matter (CRDM) framework~\cite{Bringmann:2018cvk,Ema:2018bih,Alvey:2019zaa,Ema:2020ulo,Wang:2021nbf,Cappiello:2019qsw,PROSPECT:2021awi,Maity:2022exk,Alvey:2022pad,CDEX:2022fig,Super-Kamiokande:2022ncz,Bell:2023sdq,Dutta:2024kuj,Ghosh:2024dqw,Cappiello:2024acu,Guha:2024mjr,Jeesun:2025gzt,Das:2024ghw,Guha:2025tjf,DeMarchi:2025uoo}, which offers a promising pathway to access the elusive sub-GeV regime.
This mechanism posits that relativistic cosmic rays can upscatter halo DM particles via elastic collisions, imparting sufficient kinetic energy to accelerate them to velocities approaching $\sim0.1c$.
Such acceleration enables even sub-GeV DM to deposit detectable keV-scale energies in terrestrial detectors, thereby circumventing the kinematic limitations of conventional searches.

Recently, following the release of the latest data from the LZ experiment~\cite{LZ:2024zvo}, the LZ collaboration applied the CRDM approach to derive model-independent constraints on the DM-nucleon scattering cross-section~\cite{LZ:2025iaw}, assuming a constant interaction.
While this provides a valuable baseline, it does not connect to specific particle-physics models, limiting comparisons with other search strategies and theoretical scenarios.

In this work, we build upon the model-independent LZ results by incorporating specific DM particle properties through a set of well-motivated benchmark models.
We systematically consider four types of mediator scenarios: scalar, vector, pseudoscalar, and axial-vector interactions.
For each model, we derive constraints on both the DM-nucleon scattering cross-section and the underlying particle physics parameters---most notably the product of couplings $g_\chi g_N / 4\pi $ as a function of the mediator mass. A key focus of our analysis is the behavior in the regime of extremely light mediators, where the momentum transfer dependence of the interaction significantly impacts the exclusion limits. This approach allows a more direct connection between experimental results and specific particle physics models, facilitating comparisons with other DM searches and theoretical predictions.

The paper is structured as follows: 
Sec.~\ref{model} outlines the theoretical framework of particle models, CRDM scattering and detector response modeling.
Sec.~\ref{result} presents our constraints and discusses their implications. 
Sec.~\ref{conclusion} summarizes our findings and discusses their implications.

\section{Boosted Dark Matter}
\label{model}

\subsection{Particle Model}
Considering a DM particle consisting of a Dirac fermion $\chi$ with mass $m_\chi$, we model the DM-nucleon interactions within a simplified framework, specifically investigating four mediators: a scalar $\phi$, a vector $V_\mu$, an axial-vector $A_\mu$, and a pseudoscalar $\eta$. These yield two distinct interaction types: the scalar and vector mediators lead to spin-independent (SI) scattering, and the axial and pseudoscalar mediators lead to spin-dependent (SD) scattering. The effective Lagrangians for the coupling involving these mediators, nucleons $N$, and a fermionic DM particle $\chi$ are~\cite{Bell:2023sdq}: 
\begin{align}
&\mathcal{L}_{\mathrm{int}}^{\mathrm{scalar}} = g_{\chi \phi} \, \phi \, \bar{\chi} \chi + g_{N \phi} \, \phi \, \bar{N} N, \\
&\mathcal{L}_{\mathrm{int}}^{\mathrm{vector}} = g_{\chi V} \, V^{\mu} \, \bar{\chi} \gamma_{\mu} \chi + g_{N V} \, V^{\mu} \, \bar{N} \gamma_{\mu} N, \\
&\mathcal{L}_{\mathrm{int}}^{\mathrm{axialvector}} = g_{\chi A} \, A^{\mu} \, \bar{\chi} \gamma_{\mu} \gamma^{5} \chi + g_{N A} \, A^{\mu} \, \bar{N} \gamma_{\mu} \gamma^{5} N, \\
&\mathcal{L}_{\mathrm{int}}^{\mathrm{pseudoscalar}} = g_{\chi \eta} \, \eta \, \bar{\chi} \gamma^{5} \chi + g_{N \eta} \, \eta \, \bar{N} \gamma^{5} N.
\end{align}

We adopt the isoscalar coupling framework~\cite{Ema:2020ulo}, assuming equal couplings to up and down quarks, which implies equal couplings to protons and neutrons.
This simplification leads to a more tractable form of the hadronic form factors without significantly altering the overall conclusions.

The corresponding cross sections are: 
\begin{align}
    &\left(\frac{{\rm d}\sigma_{it}}{{\rm d}T_t}\right)_{\text{scalar}} 
        = g_{t\phi}^{2} g_{i\phi}^{2} A_{i}^{2} G_{S}^{2}(q^{2}) 
        \times \frac{(2m_t + T_t)(4m_i^2 + 2m_t T_t)}{16\pi (T_i^2 + 2m_i T_i) (m_\phi^2 + 2m_t T_t)^2}, \\
    &\left(\frac{{\rm d}\sigma_{it}}{{\rm d}T_t}\right)_{\text{vector}} 
        = g_{tV}^{2} g_{iV}^{2} A_{i}^{2} G_{V}^{2}(q^{2}) 
        \times \frac{ 2m_t (m_i + T_i)^2 - \left[ (m_i + m_t)^2 + 2m_t T_i \right] T_t + m_t T_t^2 }{4\pi (T_i^2 + 2m_i T_i) (m_V^2 + 2m_t T_t)^2}, \\
    &\left(\frac{{\rm d}\sigma_{it}}{{\rm d}T_t}\right)_{\text{axial}} 
        = g_{tA}^{2} g_{iA}^{2} G_{A}^{2}(q^{2}) 
        \times \frac{ 2m_t \left( T_i^2 + 3m_i^2 - T_i T_t + \frac{T_t^2}{2} \right) + T_t (m_i - m_t)^2 }{4\pi (T_i^2 + 2m_i T_i) (m_A^2 + 2m_t T_t)^2}, \\
    &\left(\frac{{\rm d}\sigma_{it}}{{\rm d}T_t}\right)_{\text{pseudoscalar}} 
        = g_{t\eta}^{2} g_{i\eta}^{2} G_{P}^{2}(q^{2})
        \times \frac{ m_t T_t^2 }{8\pi (T_i^2 + 2m_i T_i) (m_\eta^2 + 2m_t T_t)^2},
\label{eq:crosssection}
\end{align}
where $m_\phi$, $m_V$, $m_A$, $m_\eta$ are the mediator mass and the subscripts $i$ and $t$ denote to incident and target particles.
$G$ is the hadronic elastic scattering form factor. For SI scattering, we adopt the dipole form $G(Q^{2}) = 1/{\left(1 + Q^{2} / \Lambda_{i}^{2}\right)^{2}}$~\cite{Perdrisat:2006hj}. For SD scattering, the axial situation we adopt $\Lambda_p = 1.026$ GeV~\cite{MiniBooNE:2013qnd}, and the pseudoscalar situation  $G_{\text{pseudoscalar}}(q^2) = {G_{\text{axial}}(q^2)  C_q} / ({q^2 + M_{q}^{2}})$ where $C_q = 0.9\ \mathrm{GeV^2}$ and $M_q = 0.33$ GeV~\cite{Ema:2020ulo}.
This parameterization for the pseudoscalar form factor is obtained under the assumption of equal couplings to up and down quarks.
For the scattering with nuclei heavier than helium, we employ the standard Helm nuclear form factor~\cite{Helm:1956zz} to account for the loss of coherence due to the finite nuclear size.

\subsection{Cosmic-ray Boosted Dark Matter}

Galactic DM particles possess characteristic velocities $v_\chi \sim 10^{-3}c$ as described by the Standard Halo Model~\cite{Navarro:1995iw}. 
At this speed, the kinetic energy of sub-GeV DM is insufficient to produce a detectable signal in conventional direct detection experiments.
However, cosmic-ray (CR) interactions can impart relativistic boosts to DM particles through elastic scattering~\cite{Bringmann:2018cvk}. 

Treating DM particles as approximately at rest compared to high-energy CRs, the maximum kinetic energy transfer in a CR--DM ($i$--$\chi$) collision is:
\begin{equation}
T_\chi^{\mathrm{max}} = \frac{T_i(T_i + 2m_i)}{T_i + {(m_i + m_\chi)^2}/{2m_\chi}},
\label{eq:Tchi_max}
\end{equation}
where $m_i$ and $T_i$ denote the CR mass and kinetic energy. Therefore the minimum CR energy required to produce a DM particle with kinetic energy $T_\chi$ is:
\begin{equation}
T_i^{\mathrm{min}} = \left( \frac{T_\chi}{2} - m_i \right) \left[ 1 \pm \sqrt{1 + \frac{2T_\chi(m_i + m_\chi)^2}{m_\chi(2m_i - T_\chi)^2}} \right],
\label{eq:Ti_min}
\end{equation}
with the $+$ ($-$) sign applying for $T_\chi > 2m_i$ ($T_\chi < 2m_i$).

As DM particles collide with CR in space, the differential boosted DM flux in the vicinity of Earth then follows~\cite{Bringmann:2018cvk}:
\begin{equation}
\frac{{\rm d}\Phi_\chi}{{\rm d}T_\chi} = D_{\mathrm{eff}} \frac{\rho^{\mathrm{local}}_\chi}{m_\chi} \sum_i  \int_{T_i^{\mathrm{min}}}^{\infty} {\rm d}T_i \frac{{\rm d}\sigma_{\chi i}}{{\rm d}T_\chi} \frac{{\rm d}\Phi^{\mathrm{LIS}}_i}{{\rm d}T_i} ,
\label{eq:DM_flux}
\end{equation}
where $\rho^{\rm local}_{\chi}$ is the local DM density which we take to be 0.3 $\mathrm{GeV/cm^3}$, ${\rm d}\Phi^{\text{LIS}}_i / {\rm d}T_i$ is the local interstellar spectrum, and we include contributions from CR nuclei with atomic numbers $Z\leqslant28$.  
The effective propagation distance $D_{\mathrm{eff}} \equiv (1/\rho_\chi^{\mathrm{local}}) \int {\rm d}\Omega/(4\pi) \int_{\rm los} \rho_\chi \,{\rm d}\ell$~\cite{Maity:2022exk,Dent:2020syp,Bardhan:2022bdg} encapsulates the Galactic DM distribution and CR propagation geometry.
In our analysis based on the NFW profile~\cite{Navarro:1995iw,Fermi-LAT:2012pls}, we adopt $D_{\mathrm{eff}} = 1$ or $10\,\mathrm{kpc}$.

\begin{figure}[htbp]
     \centering
     \includegraphics[width=0.8\linewidth]{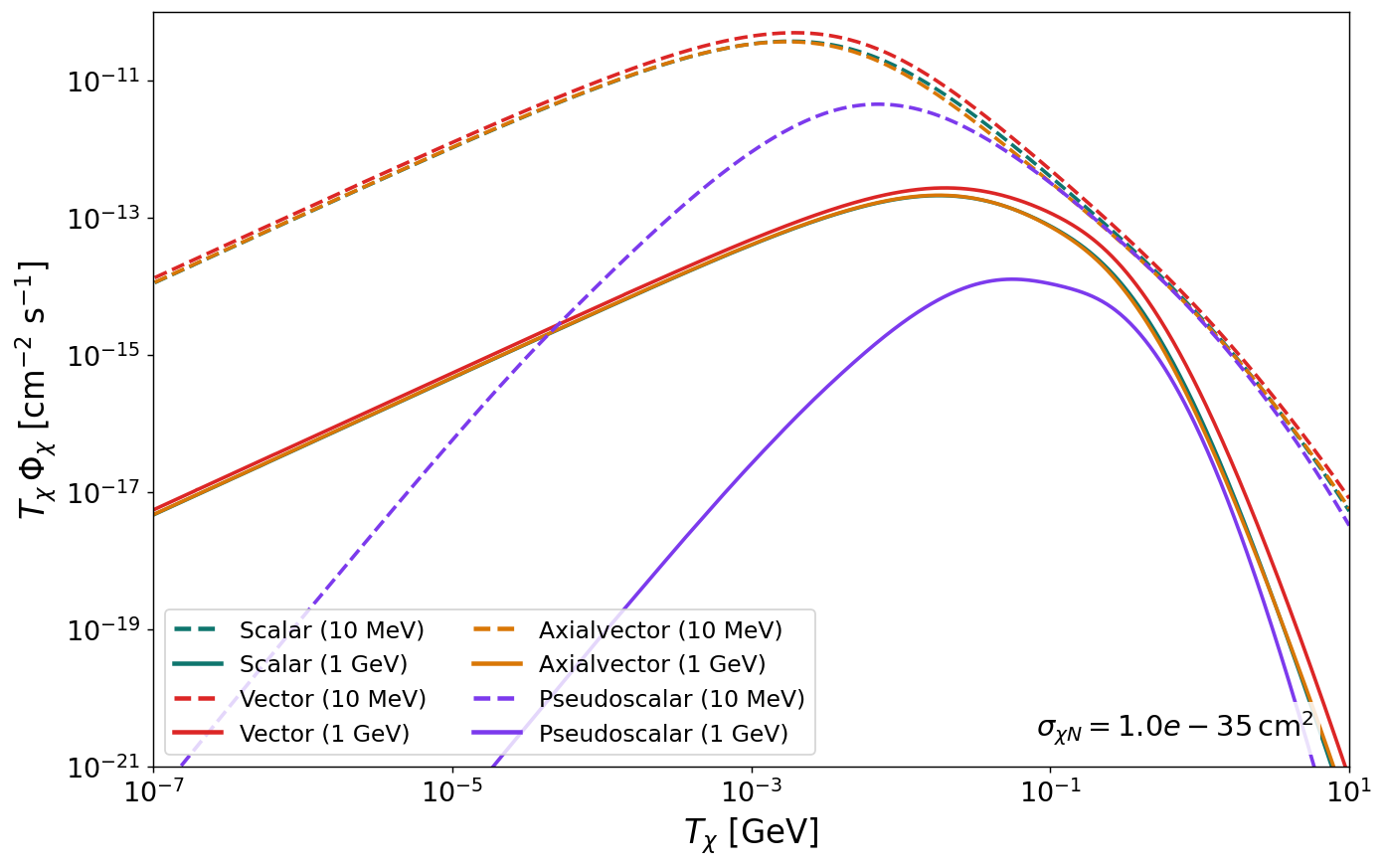}
     \caption{CRDM flux for scalar, vector, axial-vector, and pseudoscalar mediators, shown for two representative dark-matter masses $m_\chi=10$ MeV and $m_\chi=1$ GeV, with a common normalization of the scattering cross section $\sigma_{\chi N}=10^{-35}\mathrm{cm}^2$ and $D_{\mathrm{eff}} = 10\,\mathrm{kpc}$.}
     \label{flux}
\end{figure}

Fig.~\ref{flux} shows the corresponding CRDM fluxes for two representative dark-matter masses $m_\chi=10$ MeV and $m_\chi=1$ GeV across the four mediator types introduced in the previous subsection. 
Both the overall flux level and the spectral shape depend sensitively on the mediator type. 
The flux of pseudoscalar model is significantly lower than that of other models in the low-energy regime, while the other models exhibit similar behavior.

During propagation to underground detectors, DM particles undergo energy attenuation through elastic scattering with atmospheric and geological nuclei. 
The resulting energy loss is described by:
\begin{align}
\frac{{\rm d}T_{\chi}^{z}}{{\rm d}z} & = -\sum_{N}n_{N}\int_{0}^{T_{N}^{\max }}{\rm d}T_{N}\frac{{\rm d}\sigma_{\chi N}}{{\rm d}T_{N}}T_{N}
\label{eq:attenuation},
\end{align}

where $T_N$ is the nuclear recoil energy and $z$ is the depth of the laboratory.
The attenuation of DM in the overburden is obtained by numerically integrating the energy-loss equation along its propagation path through the underground. 
The dominant energy-loss mechanism is coherent elastic scattering with crustal nuclei, modeled using Helm form factors and full relativistic kinematics.
For experiments located at or near the Earth surface, the atmospheric elemental composition (N, O, and Ar) is taken into account. For underground experiments, the overburden is instead modeled as a weighted average of the most abundant crustal elements (C, O, Mg, Al, Si, K, and Ca).~\cite{Wulandari:2003cr,Alvey:2022pad}

Finally, DM particles arrive at underground laboratories and scatter off the target nuclei, generating a differential event rate given by:
\begin{align}
\frac{{\rm d}\Gamma_{N}}{{\rm d}T_{N}} = \int_{T_{\chi}^{\min}}^{\infty}{\rm d}T_{\chi}\;\frac{{\rm d}\sigma_{\chi N}}{{\rm d}T_{N}}\frac{{\rm d}\Phi_{\chi}}{{\rm d}T_{\chi}}.
\label{eq:scattering_rate}
\end{align}

By comparing this theoretically predicted event rate with experimental measurements, constraints on the relevant parameter space can be derived.

\section{Model Constraints}
\label{result}
\subsection{Constraints on Constant Cross-sections}

Constant cross-sections refers to the scenario where the differential scattering cross-section $d\sigma_{\chi N}/dT_N$ is assumed to be independent of the momentum transfer $q$. 
This is a valid approximation in the limit of a very heavy mediator (effectively a contact interaction), where the propagator simplifies to $1/m_{\text{med}}^2$. 
While both SI and SD interactions can, in principle, be considered under this assumption, we focus here on the SI case. This choice is motivated by the fact that the CR-boosting process and the subsequent detection in large-volume liquid xenon detectors like LZ and XENON1T are primarily sensitive to the coherent, SI scattering channel, which benefits from an $A^2$ (atomic mass squared) enhancement.

\begin{figure}[htbp]
     \centering
     \includegraphics[width=0.8\linewidth]{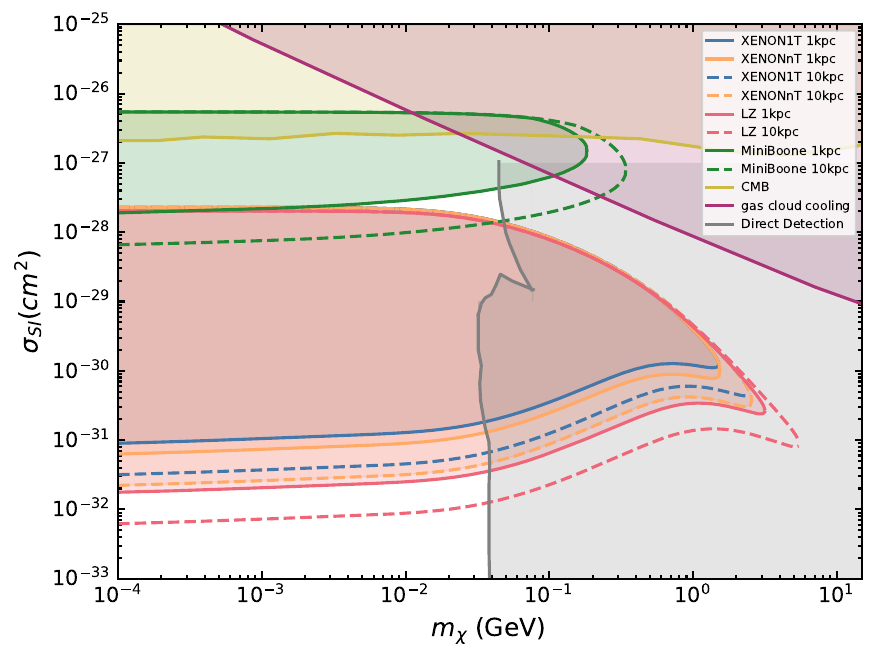}
     \caption{Bounds on SI constant DM-nucleon scattering cross-section. The LZ~\cite{LZ:2024zvo} (red), XENON1T~\cite{XENON:2018voc} (blue),XENONnT(orange)~\cite{XENON:2023cxc} and MiniBooNE~\cite{Karagiorgi:2006jf} (green) constraints are shown with the corresponding line styles, and the solid and dashed lines represent the local value out to a distance of 1~kpc and 10~kpc. We compare our result with CMB observations~\cite{Xu:2018efh} (orange), gas cloud cooling~\cite{Bhoonah:2018wmw} (purple) and direct detection limit (gray) which incorporates results from SuperCDMS~\cite{SuperCDMS:2018mne}, DarkSide-50~\cite{DarkSide:2022dhx,DarkSide-50:2022qzh}, DAMIC~\cite{DAMIC:2016lrs}, NEWS-G~\cite{NEWS-G:2017pxg}, CRESST~\cite{CRESST:2019jnq}, CDEX~\cite{CDEX:2019hzn}, XENON~\cite{XENON:2019zpr}, and EDELWEISS~\cite{EDELWEISS:2019vjv,EDELWEISS:2022ktt}. }
     \label{result2}
\end{figure}

Incorporating data from XENON1T~\cite{XENON:2018voc}, XENONnT(orange)~\cite{XENON:2023cxc}, MiniBooNE~\cite{Karagiorgi:2006jf}, and LZ~\cite{LZ:2024zvo}, the constraints on constant DM-nucleon scattering cross-sections are shown in Fig.~\ref{result2}, where the shaded areas represent the excluded parameter space.
We also include constraints from CMB observations~\cite{Xu:2018efh}, gas cloud cooling~\cite{Bhoonah:2018wmw} and other direct detection experiments~\cite{SuperCDMS:2018mne,DarkSide:2022dhx,DarkSide-50:2022qzh,DAMIC:2016lrs,NEWS-G:2017pxg,CRESST:2019jnq,CDEX:2019hzn,XENON:2019zpr,EDELWEISS:2019vjv,EDELWEISS:2022ktt}. 
While results similar to those from LZ can be found in Ref.~\cite{LZ:2025iaw}, we independently calculate the limits under the constant cross-section assumption. 
Our results are derived from published event-rate limits rather than from raw data.
Furthermore, by combining the LZ results with those from other experiments, we obtain an extended coverage of the parameter space. 
In particular, Fig.~\ref{result2} includes the constraints from MiniBooNE, which significantly enlarges the excluded region. Due to its shallow depth, MiniBooNE is sensitive to larger cross sections that would otherwise be inaccessible to deep underground experiments due to Earth attenuation.

For each experiment, the excluded region lies between its upper and lower bounding curves.
The upper boundary is primarily determined by the experiment's depth: an excessively large cross-section would lead to DM scattering in the Earth's crust before reaching the detector, resulting in insufficient event rates. 
The lower boundary implies that a cross-section too small cannot produce an excess of events. 
The XENON1T and LZ detectors are situated approximately 1,400~m underground, whereas MiniBooNE operates in a shallow subsurface facility. 
This difference in depth provides complementary coverage of the parameter space, significantly extending the excluded region.

Given the current uncertainty in $D_{\mathrm{eff}}$, which arises from the Galactic DM velocity distribution and local density fluctuations~\cite{Read:2014qva}, the final $\sigma_{\mathrm{SI}}$ constraints scale as $\sqrt{D_{\mathrm{eff}}}$~\cite{Maity:2022exk}, reflecting the geometric dependence of signal accumulation time in directional detection schemes.
In the following analysis we adopt the $D_{\mathrm{eff}}$ value of 10~kpc.

\subsection{Constraints on Light Mediator Models}

\begin{figure}[htbp]
    \centering
    \begin{subfigure}{0.48\textwidth}
        \centering
        \includegraphics[width=\linewidth]{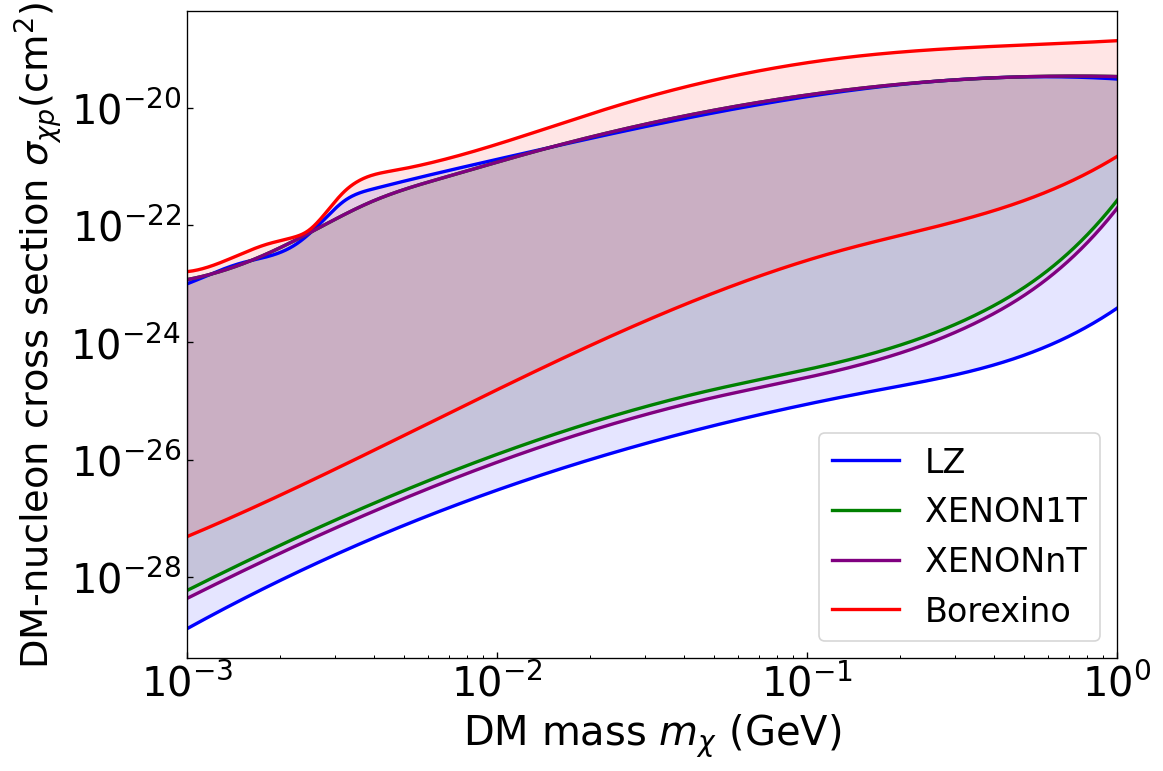}
        \caption{Scalar $1\,\rm{MeV}$}
        \label{fig:scalar1}
    \end{subfigure}
    \hfill
    \begin{subfigure}{0.48\textwidth}
        \centering
        \includegraphics[width=\linewidth]{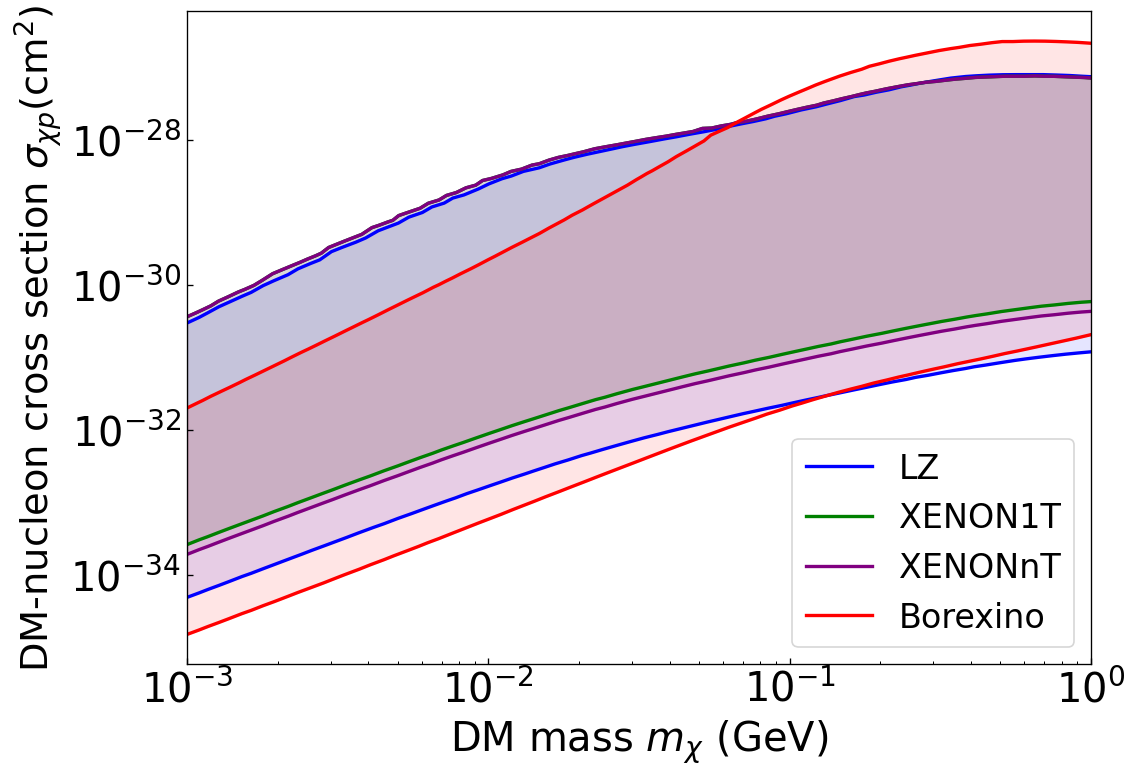}
        \caption{Scalar $1\,\rm{GeV}$}
        \label{fig:scalar1000}
    \end{subfigure}
    
    \vspace{0.5cm}
    
    \begin{subfigure}{0.48\textwidth}
        \centering
        \includegraphics[width=\linewidth]{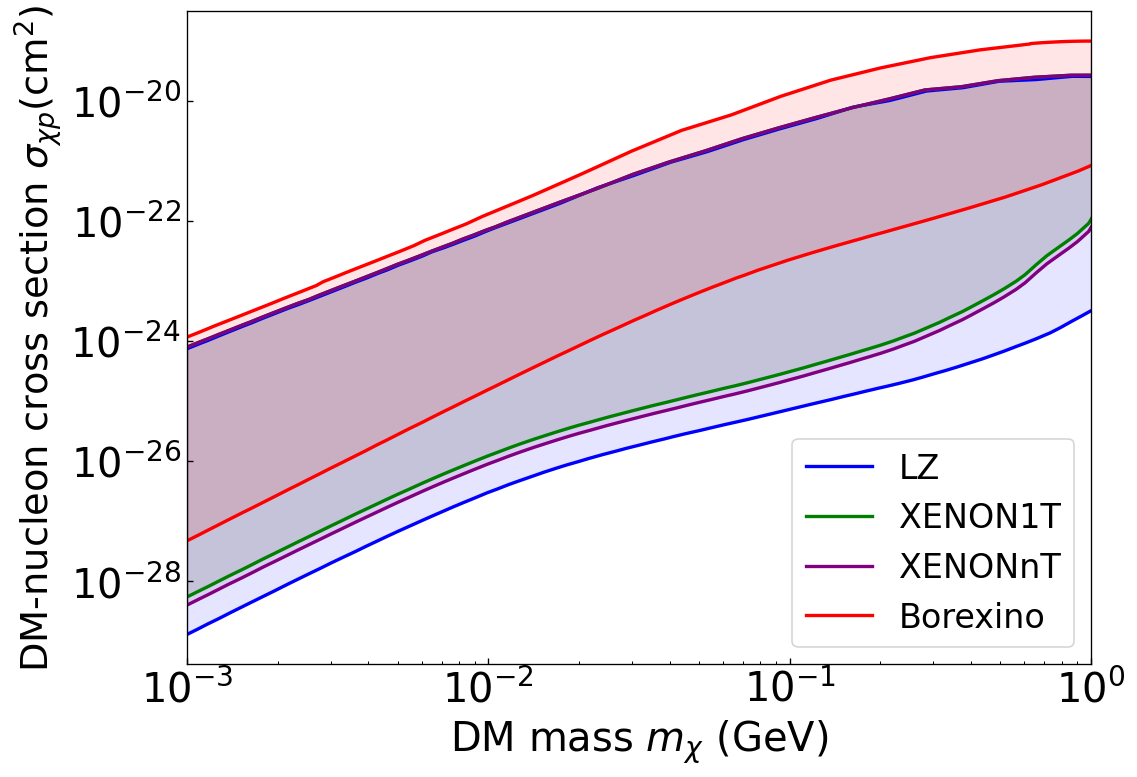}
        \caption{Vector $1\,\rm{MeV}$}
        \label{fig:vector1}
    \end{subfigure}
    \hfill
    \begin{subfigure}{0.48\textwidth}
        \centering
        \includegraphics[width=\linewidth]{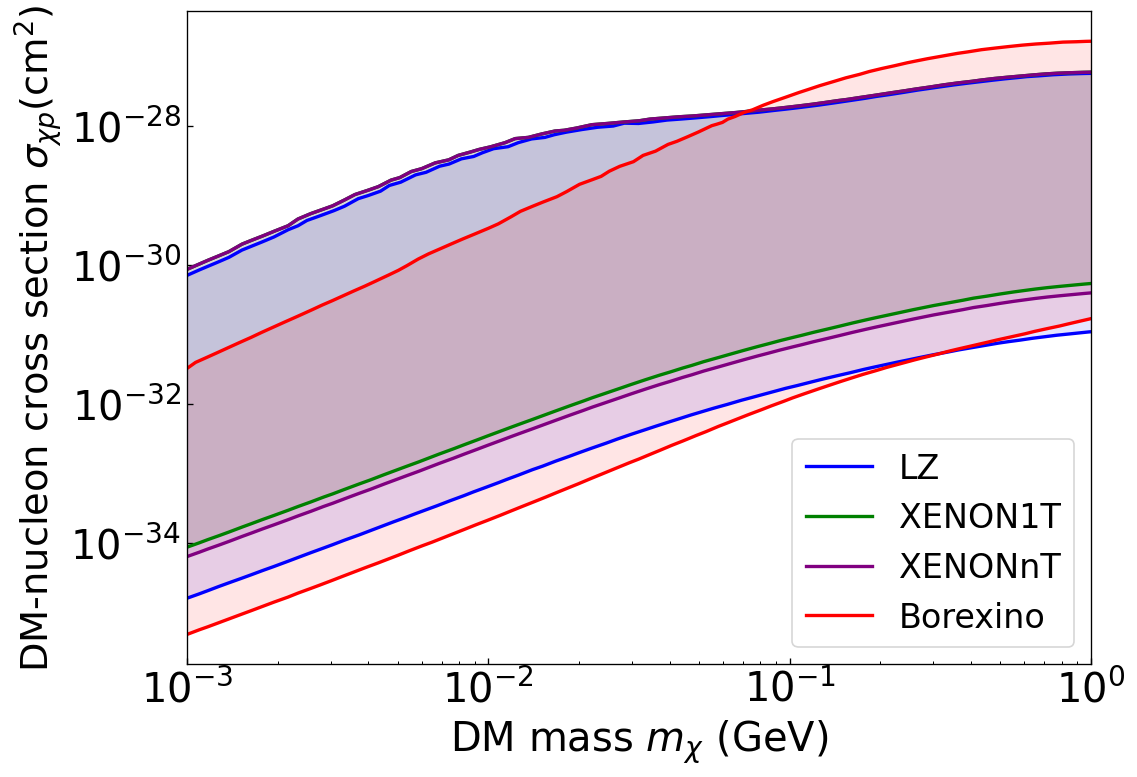}
        \caption{Vector $1\,\rm{GeV}$}
        \label{fig:vector1000}
    \end{subfigure}
    \caption{Constraints on the DM-nucleon scattering cross section for scalar and vector mediator models. Left (right) panels show the results for a mediator mass of 1\,MeV (1\,GeV). Blue, green, and red curves correspond to the LZ, XENON1T, XENONnT and Borexino limits, and the shaded regions indicate the excluded parameter space.}
    \label{fig:modelsig}
\end{figure}

Unlike the model-independent assumption of a constant scattering cross-section, light-mediator models exhibit a pronounced dependence of the DM–nucleon scattering cross-section on the momentum transfer, determined by the mediator type and mass.
In Fig.~\ref{fig:modelsig}, we present the resulting constraints for scalar and vector mediators with light (1\,MeV) and heavy (1\,GeV) mediator masses.
The $\sigma_{\chi p}$ denotes the total elastic scattering cross section between DM particle and stationary proton, obtained by integrating the corresponding differential cross section $d\sigma_{\chi p}/dT_p$. 
As the DM mass increases, the constraints systematically weaken due to the decreasing scattering cross-section.

Furthermore, the relative sensitivities of different experiments vary significantly between the light- and heavy-mediator regimes.
In this work, we directly adopt the final calibrated data products, specifically extracting the 90\% C.L. exclusion curve on the differential event rate from Ref.~\cite{LZ:2024zvo}.
For light mediators, the scattering is strongly weighted toward low momentum transfer, corresponding to very low nuclear recoil energies.
In this regime, the LZ experiment~\cite{LZ:2024zvo}, benefiting from its exceptionally low energy threshold and the strong coherence enhancement of SI interactions, provides the most stringent constraints.
For heavier mediators, the low-momentum-transfer enhancement is absent, rendering the sensitivity largely statistics-driven. As a result, neutrino experiments such as Borexino~\cite{Borexino:2013zhu,BOREXINO:2023ygs}, with their large effective target masses and exposures, become competitive with direct detection experiments.

\begin{figure}[htbp]
    \centering
    \begin{subfigure}{0.45\textwidth}
        \includegraphics[width=\linewidth]{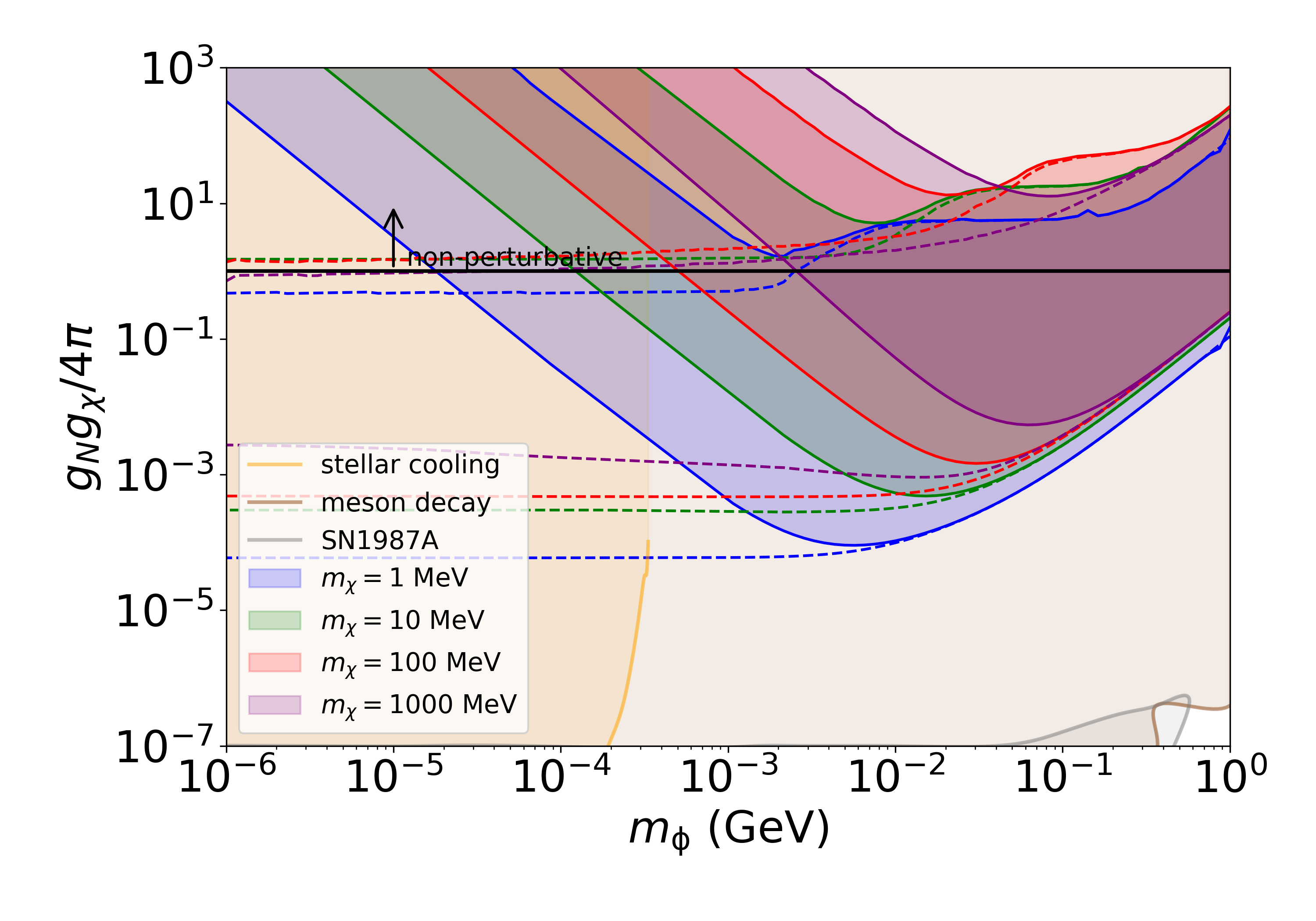}
        \caption{Scalar}
        \label{fig:para_scalar}
    \end{subfigure}
    \begin{subfigure}{0.45\textwidth}
        \includegraphics[width=\linewidth]{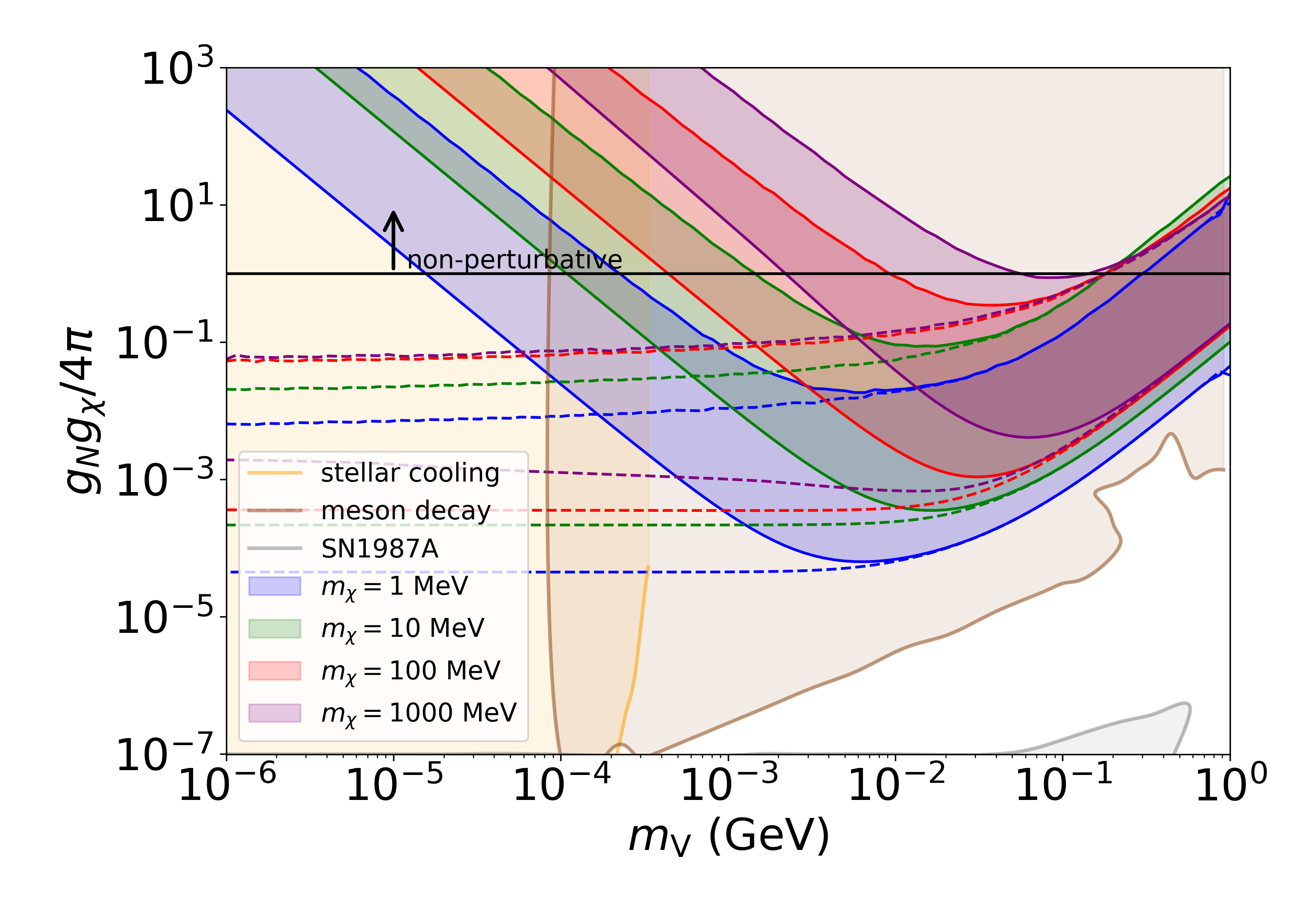}
        \caption{Vector}
        \label{fig:para_vector}
    \end{subfigure}
    \begin{subfigure}{0.45\textwidth}
        \includegraphics[width=\linewidth]{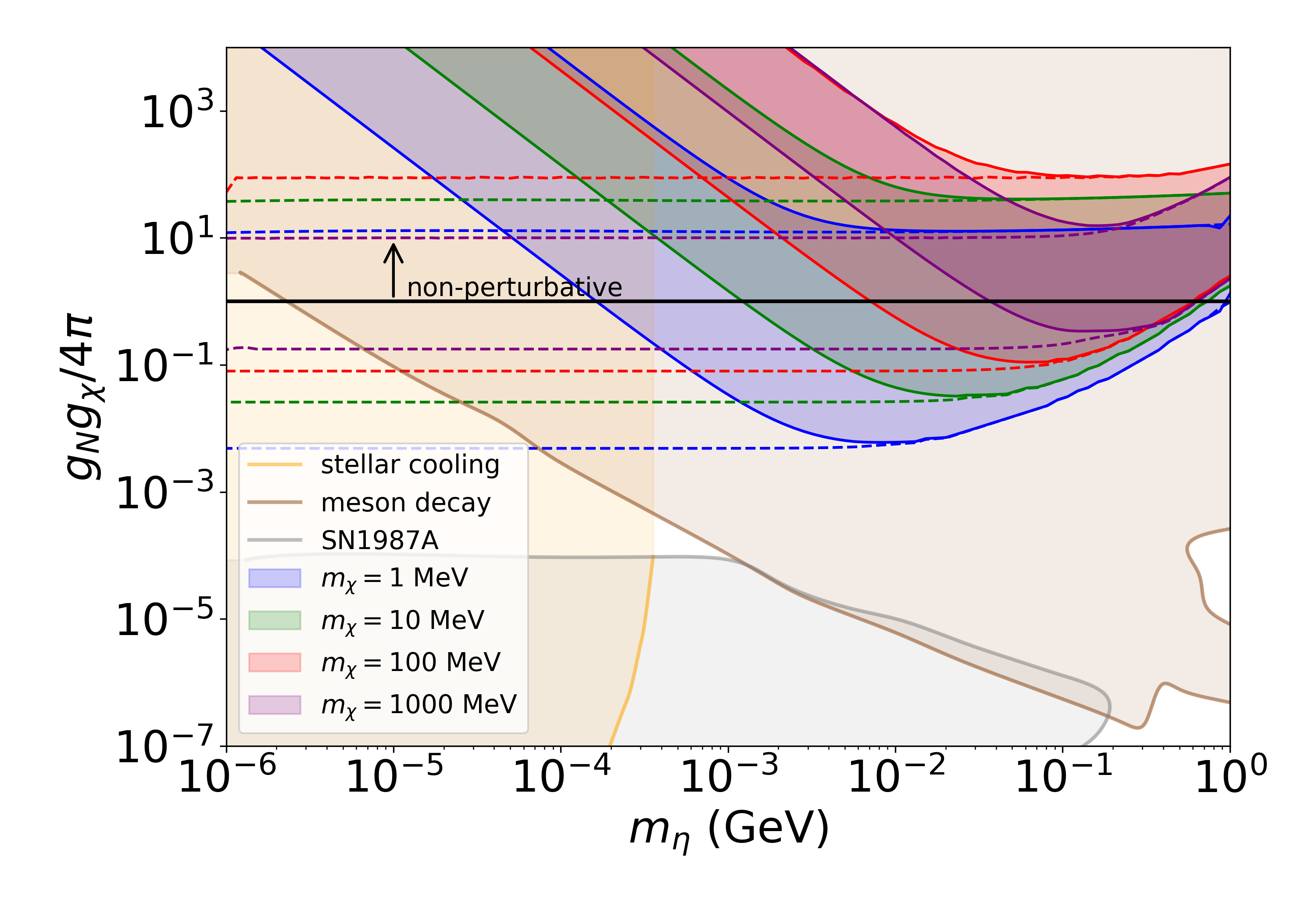}
        \caption{Pseudoscalar}
        \label{fig:para_pseudoscalar}
    \end{subfigure}
    \begin{subfigure}{0.45\textwidth}
        \includegraphics[width=\linewidth]{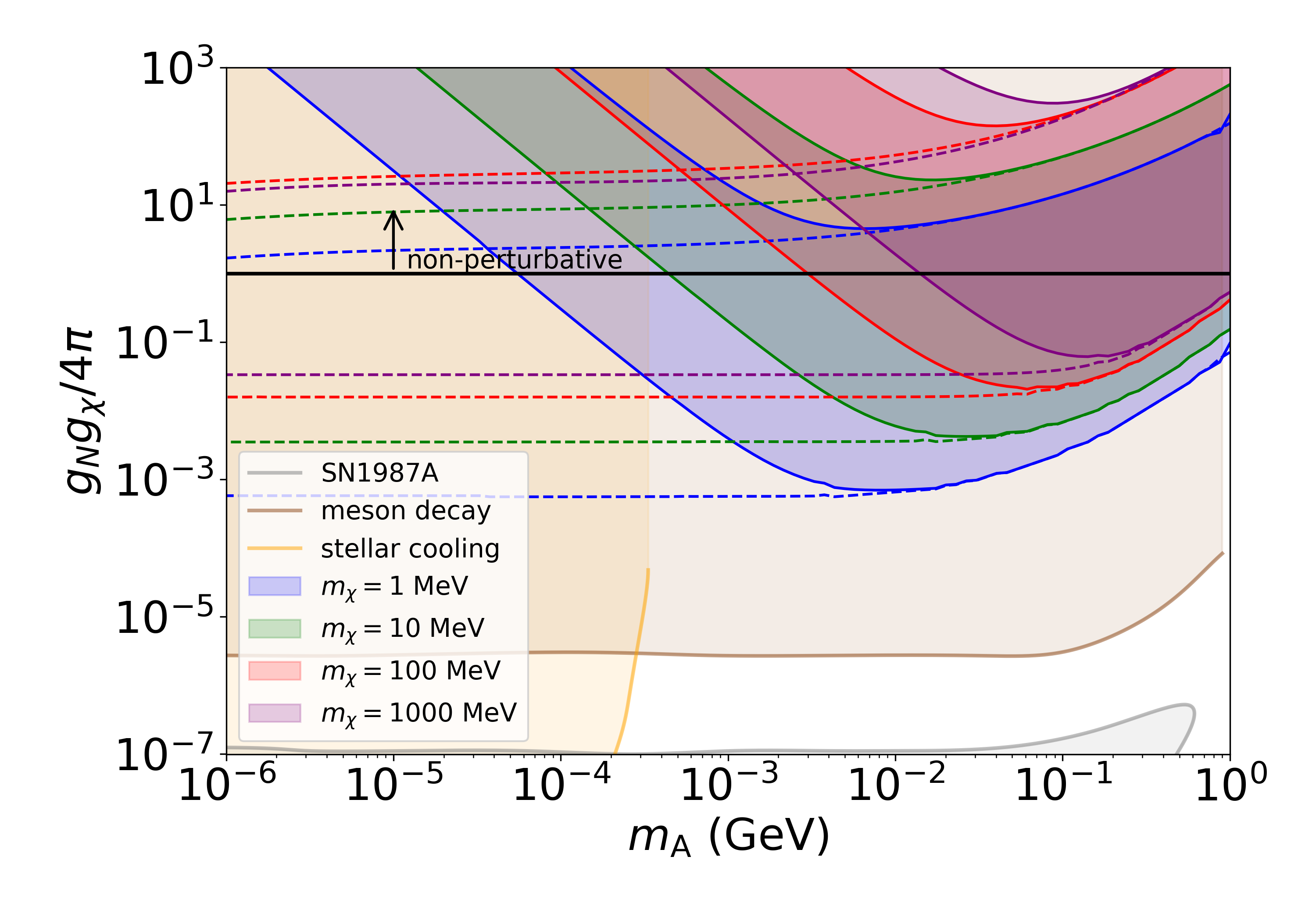}
        \caption{Axial-vector}
        \label{fig:para_axialvector}
    \end{subfigure}
    \caption{Constraints on DM-nucleon couplings for the four benchmark DM masses.
    The upper panels show SI results for scalar and vector mediators obtained from the LZ experiment, while the lower panels show SD results for pseudoscalar and axial-vector mediators from the Borexino experiment.
    Blue, green, red, and purple lines correspond to DM masses of 1~MeV, 10~MeV, 100~MeV, and 1000~MeV.
    Solid lines indicate the full calculation including both the mediator mass and momentum terms in the propagator, while dashed lines retain only the mass term.
    For comparison, we indicate the regions where existing constraints already exclude the new light mediator.}
        
    \label{fig:SI-SD}
\end{figure}

We further study the four benchmark models introduced in Sec.~\ref{model} and derive experimental constraints on the coupling product $g_N g_\chi/4\pi$.
Scalar and vector mediators induce SI scattering, whereas pseudoscalar and axial-vector mediators give rise to SD interactions.
SI constraints are primarily obtained from the LZ experiment, while SD constraints are mostly provided by hydrogen-target neutrino experiments such as Borexino. 
The resulting exclusion regions for four benchmark DM masses are shown in Fig.~\ref{fig:SI-SD}.
For comparison, Fig.~\ref{fig:SI-SD} also includes existing constraints on the mediator from meson decays~\cite{Knapen:2017xzo,Dror:2017ehi,Dolan:2014ska,Kahn:2016vjr}, stellar cooling~\cite{Hardy:2016kme}, and the trapping and cooling limits derived from SN1987A~\cite{Rrapaj:2015wgs}. These bounds generally apply to the mediator coupling to Standard Model particles. 
Consequently, recasting them as limits on $g_N g_\chi/4\pi$ requires additional assumptions about the dark matter coupling. 
Throughout this work, we adopt the assumptions in Ref.~\cite{Bell:2023sdq}.

In the light-mediator regime, $m_{\rm med}\lesssim 10~\mathrm{MeV}$, the propagator 
gives rise to an approximate scaling of the differential cross section ${\rm d}\sigma / {\rm d}E_R \propto 1/{q^4}$, where $q$ is the momentum transfer. This strongly enhances small-$q$ scattering and shifts the recoil spectrum towards lower nuclear recoil energies. Consequently, the experimental sensitivity becomes highly dependent on the detector energy threshold, leading to a rapid change of the exclusion limits.

The solid and dashed curves represent two theoretical prescriptions for translating the excluded scattering cross section into constraints on the underlying couplings. The dashed curves use the contact-interaction approximation, in which the momentum dependence of the mediator propagator is neglected and the squared amplitude is assumed to scale as $m_{\rm med}^{-4}$. This treatment is reliable only in the heavy-mediator limit, $m_{\rm med}^2 \gg q_{\rm typ}^2$, where $q_{\rm typ}\simeq\sqrt{2m_T E_R^{\rm typ}}$ denotes the characteristic momentum transfer. In contrast, the solid curves retain the full propagator factor $(m_{\rm med}^2+q^2)^{-2}$ and therefore account for finite-momentum-transfer effects. The two predictions coincide for $m_{\rm med}\gg q_{\rm typ}$, while they deviate increasingly for $m_{\rm med}\lesssim q_{\rm typ}$, where the contact-interaction approximation breaks down.

\section{Conclusion}
\label{conclusion}
CRDM provides a unique opportunity to probe sub-GeV DM, producing detectable nuclear recoil signals in underground detectors that are otherwise inaccessible to conventional direct detection experiments.
We consider both model-independent and model-dependent DM–nucleon scattering cross-sections and derive updated constraints using data from LZ, XENON, and Borexino.
In particular, we provide limits on the DM–nucleon coupling for four benchmark mediator models (scalar, vector, pseudoscalar, and axial-vector), significantly extending the reach of terrestrial searches into the sub-GeV mass regime.

Our results reveal a characteristic turnover in the exclusion limits for mediator masses around $10^{-2}$–$10^{-3}\,\mathrm{GeV}$, corresponding to the transition between momentum-transfer dominated and mass dominated scattering.
This feature illustrates the significant impact of momentum dependence in light-mediator scenarios, which must be taken into account when interpreting direct detection constraints.
These findings demonstrate that CRDM provides a viable approach to probing sub-GeV DM, extending the sensitivity of current underground experiments.

\section*{Acknowledgments}
This work is supported by the National Key R\&D Program of China (Grant No. 2022YFF0503304), the National Natural Science Foundation of China (Grant Nos. 12373002, 12220101003, 11773075), the Youth Innovation Promotion Association of Chinese Academy of Sciences (Grant No. 2016288), and the Jiangsu Province Post Doctoral Foundation (Grant No. 2024ZB713).



\begin{thebibliography}{99}
\bibitem{Zwicky:1933gu}
F.~Zwicky,
Helv. Phys. Acta \textbf{6} (1933), 110-127

\bibitem{Planck:2018vyg}
N.~Aghanim \textit{et al.} [Planck],
Astron. Astrophys. \textbf{641} (2020), A6
[erratum: Astron. Astrophys. \textbf{652} (2021), C4]
[arXiv:1807.06209 [astro-ph.CO]].

\bibitem{Rubin:1980zd}
V.~C.~Rubin, N.~Thonnard and W.~K.~Ford, Jr.,
Astrophys. J. \textbf{238} (1980), 471

\bibitem{Clowe:2006eq}
D.~Clowe, M.~Bradac, A.~H.~Gonzalez, M.~Markevitch, S.~W.~Randall, C.~Jones and D.~Zaritsky,
Astrophys. J. Lett. \textbf{648} (2006), L109-L113
[arXiv:astro-ph/0608407 [astro-ph]].

\bibitem{Bertone:2004pz}
G.~Bertone, D.~Hooper and J.~Silk,
Phys. Rept. \textbf{405} (2005), 279-390
[arXiv:hep-ph/0404175 [hep-ph]].

\bibitem{Lewin:1995rx}
J.~D.~Lewin and P.~F.~Smith,
Astropart. Phys. \textbf{6} (1996), 87-112

\bibitem{XENON:2019zpr}
E.~Aprile \textit{et al.} [XENON],
Phys. Rev. Lett. \textbf{123} (2019) no.24, 241803
[arXiv:1907.12771 [hep-ex]].

\bibitem{XENON:2022ltv}
E.~Aprile \textit{et al.} [XENON],
Phys. Rev. Lett. \textbf{129} (2022) no.16, 161805
[arXiv:2207.11330 [hep-ex]].

\bibitem{XENON:2023cxc}
E.~Aprile \textit{et al.} [XENON],
Phys. Rev. Lett. \textbf{131} (2023) no.4, 041003
[arXiv:2303.14729 [hep-ex]].

\bibitem{LZ:2022lsv}
J.~Aalbers \textit{et al.} [LZ],
Phys. Rev. Lett. \textbf{131} (2023) no.4, 041002
[arXiv:2207.03764 [hep-ex]].

\bibitem{LZ:2023poo}
J.~Aalbers \textit{et al.} [LZ],
Phys. Rev. D \textbf{108} (2023) no.7, 072006
[arXiv:2307.15753 [hep-ex]].

\bibitem{LZ:2024zvo}
J.~Aalbers \textit{et al.} [LZ],
Phys. Rev. Lett. \textbf{135} (2025) no.1, 011802
[arXiv:2410.17036 [hep-ex]].

\bibitem{PandaX-II:2020oim}
Q.~Wang \textit{et al.} [PandaX-II],
Chin. Phys. C \textbf{44} (2020) no.12, 125001
[arXiv:2007.15469 [astro-ph.CO]].

\bibitem{PandaX:2024qfu}
Z.~Bo \textit{et al.} [PandaX],
Phys. Rev. Lett. \textbf{134} (2025) no.1, 011805
[arXiv:2408.00664 [hep-ex]].

\bibitem{PandaX:2025rrz}
M.~Zhang \textit{et al.} [PandaX],
Phys. Rev. Lett. \textbf{135} (2025) no.21, 211001
[arXiv:2507.11930 [hep-ex]].

\bibitem{DEAP:2019yzn}
R.~Ajaj \textit{et al.} [DEAP],
Phys. Rev. D \textbf{100} (2019) no.2, 022004
[arXiv:1902.04048 [astro-ph.CO]].

\bibitem{DarkSide:2018kuk}
P.~Agnes \textit{et al.} [DarkSide],
Phys. Rev. D \textbf{98} (2018) no.10, 102006
[arXiv:1802.07198 [astro-ph.CO]].

\bibitem{Knapen:2017xzo}
S.~Knapen, T.~Lin and K.~M.~Zurek,
Phys. Rev. D \textbf{96} (2017) no.11, 115021
[arXiv:1709.07882 [hep-ph]].

\bibitem{Essig:2011nj}
R.~Essig, J.~Mardon and T.~Volansky,
Phys. Rev. D \textbf{85} (2012), 076007
[arXiv:1108.5383 [hep-ph]].

\bibitem{SuperCDMS:2018mne}
R.~Agnese \textit{et al.} [SuperCDMS],
Phys. Rev. Lett. \textbf{121} (2018) no.5, 051301
[erratum: Phys. Rev. Lett. \textbf{122} (2019) no.6, 069901]
[arXiv:1804.10697 [hep-ex]].

\bibitem{DarkSide:2022dhx}
P.~Agnes \textit{et al.} [DarkSide],
Phys. Rev. Lett. \textbf{130} (2023) no.10, 10
[arXiv:2207.11967 [hep-ex]].

\bibitem{DarkSide-50:2022qzh}
P.~Agnes \textit{et al.} [DarkSide-50],
Phys. Rev. D \textbf{107} (2023) no.6, 6
[arXiv:2207.11966 [hep-ex]].

\bibitem{DAMIC:2016lrs}
A.~Aguilar-Arevalo \textit{et al.} [DAMIC],
Phys. Rev. D \textbf{94} (2016) no.8, 082006
[arXiv:1607.07410 [astro-ph.CO]].

\bibitem{NEWS-G:2017pxg}
Q.~Arnaud \textit{et al.} [NEWS-G],
Astropart. Phys. \textbf{97} (2018), 54-62
[arXiv:1706.04934 [astro-ph.IM]].

\bibitem{CRESST:2019jnq}
A.~H.~Abdelhameed \textit{et al.} [CRESST],
Phys. Rev. D \textbf{100} (2019) no.10, 102002
[arXiv:1904.00498 [astro-ph.CO]].

\bibitem{CDEX:2019hzn}
Z.~Z.~Liu \textit{et al.} [CDEX],
Phys. Rev. Lett. \textbf{123} (2019) no.16, 161301
[arXiv:1905.00354 [hep-ex]].

\bibitem{EDELWEISS:2019vjv}
E.~Armengaud \textit{et al.} [EDELWEISS],
Phys. Rev. D \textbf{99} (2019) no.8, 082003
[arXiv:1901.03588 [astro-ph.GA]].

\bibitem{EDELWEISS:2022ktt}
E.~Armengaud \textit{et al.} [EDELWEISS],
Phys. Rev. D \textbf{106} (2022) no.6, 062004
[arXiv:2203.03993 [astro-ph.GA]].

\bibitem{Essig:2017kqs}
R.~Essig, T.~Volansky and T.~T.~Yu,
Phys. Rev. D \textbf{96} (2017) no.4, 043017
[arXiv:1703.00910 [hep-ph]].

\bibitem{Chao:2021liw}
W.~Chao, M.~Jin and Y.~Q.~Peng,
Phys. Rev. D \textbf{107} (2023) no.9, 093009
[arXiv:2109.14944 [hep-ph]].

\bibitem{Wang:2025jhy}
G.~Wang, B.~Y.~Su, L.~Zu and L.~Feng,
Eur. Phys. J. C \textbf{85} (2025) no.11, 1348
[arXiv:2503.22148 [astro-ph.HE]].

\bibitem{Zhang:2024qof}
B.~Zhang, C.~B.~Luo and L.~Feng,
Phys. Rev. D \textbf{111} (2025) no.8, 083006
[arXiv:2412.00470 [astro-ph.HE]].

\bibitem{Guo:2023kqt}
J.~Guo, L.~Wu and B.~Zhu,
Phys. Lett. B \textbf{840} (2023), 137853
[arXiv:2302.06159 [hep-ph]].

\bibitem{Tang:2025vqf}
T.~P.~Tang, M.~Yang, K.~K.~Duan, Y.~L.~S.~Tsai and Y.~Z.~Fan,
JCAP \textbf{11} (2025), 013
[arXiv:2505.05359 [hep-ph]].

\bibitem{Liang:2024xcx}
Z.~L.~Liang, L.~Su, L.~Wu and B.~Zhu,
Phys. Rev. Lett. \textbf{134} (2025) no.7, 071001
[arXiv:2401.11971 [hep-ph]].

\bibitem{Su:2023zgr}
L.~Su, L.~Wu and B.~Zhu,
Sci. China Phys. Mech. Astron. \textbf{67} (2024) no.2, 221012
[arXiv:2308.02204 [hep-ph]].

\bibitem{Chen:2024njd}
Y.~T.~Chen, S.~Matsumoto, T.~P.~Tang, Y.~L.~S.~Tsai and L.~Wu,
JHEP \textbf{05} (2024), 281
[arXiv:2403.02721 [hep-ph]].

\bibitem{Zhang:2024mmg}
C.~Zhang, L.~Zu, H.~Z.~Chen, Y.~L.~S.~Tsai and Y.~Z.~Fan,
JCAP \textbf{08} (2024), 003
[arXiv:2402.18880 [astro-ph.CO]].

\bibitem{Wang:2025tdx}
Y.~N.~Wang, X.~C.~Duan, T.~P.~Tang, Z.~Wang and Y.~L.~S.~Tsai,
JCAP \textbf{08} (2025), 059
[arXiv:2502.18263 [hep-ph]].

\bibitem{Yu:2024xsy}
Y.~Yu, T.~P.~Tang and L.~Feng,
Nucl. Phys. B \textbf{1015} (2025), 116910
[arXiv:2410.21089 [hep-ph]].

\bibitem{Bringmann:2018cvk}
T.~Bringmann and M.~Pospelov,
Phys. Rev. Lett. \textbf{122} (2019) no.17, 171801
[arXiv:1810.10543 [hep-ph]].

\bibitem{Ema:2018bih}
Y.~Ema, F.~Sala and R.~Sato,
Phys. Rev. Lett. \textbf{122} (2019) no.18, 181802
[arXiv:1811.00520 [hep-ph]].

\bibitem{Alvey:2019zaa}
J.~Alvey, M.~Campos, M.~Fairbairn and T.~You,
Phys. Rev. Lett. \textbf{123} (2019), 261802
[arXiv:1905.05776 [hep-ph]].

\bibitem{Ema:2020ulo}
Y.~Ema, F.~Sala and R.~Sato,
SciPost Phys. \textbf{10} (2021) no.3, 072
[arXiv:2011.01939 [hep-ph]].

\bibitem{Wang:2021nbf}
W.~Wang, L.~Wu, W.~N.~Yang and B.~Zhu,
Phys. Rev. D \textbf{107} (2023) no.7, 073002
[arXiv:2111.04000 [hep-ph]].

\bibitem{Cappiello:2019qsw}
C.~V.~Cappiello and J.~F.~Beacom,
Phys. Rev. D \textbf{100} (2019) no.10, 103011
[erratum: Phys. Rev. D \textbf{104} (2021) no.6, 069901]
[arXiv:1906.11283 [hep-ph]].

\bibitem{PROSPECT:2021awi}
M.~Andriamirado \textit{et al.} [PROSPECT],
Phys. Rev. D \textbf{104} (2021) no.1, 012009
[arXiv:2104.11219 [hep-ex]].

\bibitem{Maity:2022exk}
T.~N.~Maity and R.~Laha,
Eur. Phys. J. C \textbf{84} (2024) no.2, 117
[arXiv:2210.01815 [hep-ph]].

\bibitem{Alvey:2022pad}
J.~Alvey, T.~Bringmann and H.~Kolesova,
JHEP \textbf{01} (2023), 123
[arXiv:2209.03360 [hep-ph]].

\bibitem{CDEX:2022fig}
R.~Xu \textit{et al.} [CDEX],
Phys. Rev. D \textbf{106} (2022) no.5, 052008
[arXiv:2201.01704 [hep-ex]].

\bibitem{Super-Kamiokande:2022ncz}
K.~Abe \textit{et al.} [Super-Kamiokande],
Phys. Rev. Lett. \textbf{130} (2023) no.3, 031802
[erratum: Phys. Rev. Lett. \textbf{131} (2023) no.15, 159903]
[arXiv:2209.14968 [hep-ex]].

\bibitem{Bell:2023sdq}
N.~F.~Bell, J.~L.~Newstead and I.~Shaukat-Ali,
Phys. Rev. D \textbf{109} (2024) no.6, 063034
[arXiv:2309.11003 [hep-ph]].


\bibitem{DeMarchi:2025uoo}
A.~G.~De Marchi, A.~Granelli, J.~Nava and F.~Sala,
JHEP \textbf{12} (2025), 136
[arXiv:2507.12278 [hep-ph]].

\bibitem{Dutta:2024kuj}
B.~Dutta, W.~C.~Huang, D.~Kim, J.~L.~Newstead, J.~C.~Park and I.~S.~Ali,
Phys. Rev. Lett. \textbf{133} (2024) no.16, 161801
[arXiv:2402.04184 [hep-ph]].

\bibitem{Ghosh:2024dqw}
D.~K.~Ghosh, T.~Gupta, M.~Heikinheimo, K.~Huitu and S.~Jeesun,
Phys. Rev. D \textbf{111} (2025) no.6, 063019
[arXiv:2411.11973 [hep-ph]].

\bibitem{Cappiello:2024acu}
C.~V.~Cappiello, Q.~Liu, G.~Mohlabeng and A.~C.~Vincent,
Phys. Rev. D \textbf{110} (2024) no.9, 095031
[arXiv:2405.00086 [hep-ph]].

\bibitem{Guha:2024mjr}
A.~Guha and J.~C.~Park,
JCAP \textbf{07} (2024), 074
[arXiv:2401.07750 [hep-ph]].


\bibitem{Jeesun:2025gzt}
S.~Jeesun,
Phys. Rev. D \textbf{111} (2025) no.10, 103022
[arXiv:2501.11569 [hep-ph]].


\bibitem{Das:2024ghw}
A.~Das, T.~Herbermann, M.~Sen and V.~Takhistov,
JCAP \textbf{07} (2024), 045
[arXiv:2403.15367 [hep-ph]].


\bibitem{Guha:2025tjf}
A.~Guha and J.~C.~Park,
[arXiv:2509.02301 [hep-ph]].



\bibitem{LZ:2025iaw}
J.~Aalbers \textit{et al.} [LZ],
Phys. Rev. Lett. \textbf{134} (2025) no.24, 241801
[arXiv:2503.18158 [hep-ex]].

\bibitem{Perdrisat:2006hj}
C.~F.~Perdrisat, V.~Punjabi and M.~Vanderhaeghen,
Prog. Part. Nucl. Phys. \textbf{59} (2007), 694-764
[arXiv:hep-ph/0612014 [hep-ph]].

\bibitem{MiniBooNE:2013qnd}
A.~A.~Aguilar-Arevalo \textit{et al.} [MiniBooNE],
Phys. Rev. D \textbf{88} (2013) no.3, 032001
[arXiv:1301.7067 [hep-ex]].


\bibitem{Helm:1956zz}
R.~H.~Helm,
Phys. Rev. \textbf{104} (1956), 1466-1475

\bibitem{Navarro:1995iw}
J.~F.~Navarro, C.~S.~Frenk and S.~D.~M.~White,
Astrophys. J. \textbf{462} (1996), 563-575
[arXiv:astro-ph/9508025 [astro-ph]].

\bibitem{Dent:2020syp}
J.~B.~Dent, B.~Dutta, J.~L.~Newstead, I.~M.~Shoemaker and N.~T.~Arellano,
Phys. Rev. D \textbf{103} (2021), 095015
[arXiv:2010.09749 [hep-ph]].

\bibitem{Bardhan:2022bdg}
D.~Bardhan, S.~Bhowmick, D.~Ghosh, A.~Guha and D.~Sachdeva,
Phys. Rev. D \textbf{107} (2023) no.1, 015010
[arXiv:2208.09405 [hep-ph]].

\bibitem{Fermi-LAT:2012pls}
M.~Ackermann \textit{et al.} [Fermi-LAT],
Astrophys. J. \textbf{761} (2012), 91
[arXiv:1205.6474 [astro-ph.CO]].

\bibitem{Wulandari:2003cr}
H.~Wulandari, J.~Jochum, W.~Rau and F.~von Feilitzsch,
Astropart. Phys. \textbf{22} (2004), 313-322
[arXiv:hep-ex/0312050 [hep-ex]].


\bibitem{XENON:2018voc}
E.~Aprile \textit{et al.} [XENON],
Phys. Rev. Lett. \textbf{121} (2018) no.11, 111302
[arXiv:1805.12562 [astro-ph.CO]].

\bibitem{Karagiorgi:2006jf}
G.~Karagiorgi, A.~Aguilar-Arevalo, J.~M.~Conrad, M.~H.~Shaevitz, K.~Whisnant, M.~Sorel and V.~Barger,
Phys. Rev. D \textbf{75} (2007), 013011
[erratum: Phys. Rev. D \textbf{80} (2009), 099902]
[arXiv:hep-ph/0609177 [hep-ph]].

\bibitem{Xu:2018efh}
W.~L.~Xu, C.~Dvorkin and A.~Chael,
Phys. Rev. D \textbf{97} (2018) no.10, 103530
[arXiv:1802.06788 [astro-ph.CO]].

\bibitem{Bhoonah:2018wmw}
A.~Bhoonah, J.~Bramante, F.~Elahi and S.~Schon,
Phys. Rev. Lett. \textbf{121} (2018) no.13, 131101
[arXiv:1806.06857 [hep-ph]].

\bibitem{Read:2014qva}
J.~I.~Read,
J. Phys. G \textbf{41} (2014), 063101
[arXiv:1404.1938 [astro-ph.GA]].

\bibitem{Borexino:2013zhu}
G.~Bellini \textit{et al.} [Borexino],
Phys. Rev. D \textbf{89} (2014) no.11, 112007
[arXiv:1308.0443 [hep-ex]].

\bibitem{BOREXINO:2023ygs}
D.~Basilico \textit{et al.} [BOREXINO],
Phys. Rev. D \textbf{108} (2023) no.10, 102005
[arXiv:2307.14636 [hep-ex]].

\bibitem{Dror:2017ehi}
J.~A.~Dror, R.~Lasenby and M.~Pospelov,
Phys. Rev. Lett. \textbf{119} (2017) no.14, 141803
[arXiv:1705.06726 [hep-ph]].

\bibitem{Dolan:2014ska}
M.~J.~Dolan, F.~Kahlhoefer, C.~McCabe and K.~Schmidt-Hoberg,
JHEP \textbf{03} (2015), 171
[erratum: JHEP \textbf{07} (2015), 103]
[arXiv:1412.5174 [hep-ph]].

\bibitem{Kahn:2016vjr}
Y.~Kahn, G.~Krnjaic, S.~Mishra-Sharma and T.~M.~P.~Tait,
JHEP \textbf{05} (2017), 002
[arXiv:1609.09072 [hep-ph]].

\bibitem{Hardy:2016kme}
E.~Hardy and R.~Lasenby,
JHEP \textbf{02} (2017), 033
[arXiv:1611.05852 [hep-ph]].

\bibitem{Rrapaj:2015wgs}
E.~Rrapaj and S.~Reddy,
Phys. Rev. C \textbf{94} (2016) no.4, 045805
[arXiv:1511.09136 [nucl-th]].


\end{thebibliography}
\end{document}